\documentclass[12pt]{article}

\usepackage{graphicx}
\usepackage{qcmc06}

\usepackage{amsmath,amsthm,amsfonts, latexsym}

\newcommand{\mtx}[2]{\left(\begin{array}{#1}#2\end{array}\right)}

\begin{document}

\pagestyle{empty}

\title{DISCRETE PHASE SPACE AND MINIMUM-UNCERTAINTY STATES}

\author{William K.~Wootters and Daniel M.~Sussman}

\affiliation{Department of Physics, Williams College\\
	     Williamstown, MA 01267, USA}

\begin{abstract}

The quantum state of a system of $n$ qubits can be represented by a Wigner 
function on a discrete phase space, each axis of the phase space taking values in
the finite field ${\mathbb F}_{2^n}$.  Within this framework, we show that one can make sense
of the notion of a ``rotationally invariant state'' of any collection of qubits, and that any such state is, in a well defined sense, a
state of minimum uncertainty.  

\end{abstract}

\setcounter{section}{0}

\section{Introduction}

A quantum state cannot be squeezed down to a point in phase space.  But there are quantum
states that closely approximate classical states, such as the coherent states of
a harmonic oscillator.  One characterization of the coherent states is based on 
the Wigner function: they are the only states for which the Wigner function is both strictly positive and
rotationally symmetric around its center (here we assume a specific
scaling of the axes appropriate for the given oscillator).  

One can also express the quantum mechanics of {\em discrete} systems in terms of phase space.
In this paper we consider a system of $n$ qubits described in
the framework of 
Ref.~\cite{GHW}, in which the discrete phase space can be pictured as a $2^n \times 2^n$
array of points.  In this framework, the discrete Wigner function preserves the
tomographic feature of the usual Wigner function, but the points of the discrete phase
space are defined abstractly and
do not come with an immediate physical interpretation.  As in the continuous case, a 
point in discrete phase space is {\em illegal} as a quantum state: it holds too much information.
But one can ask whether
there are quantum states that, like coherent states, approximate a phase-space point as
closely as possible.  We would like to identify such states and thereby to give 
more physical meaning to the discrete phase space.  In this paper we focus primarily on the
second of the two properties mentioned above: invariance under rotations.   We will see that
one can make sense of this notion in the discrete space and that rotationally
invariant states exist for any number of qubits.   

The most interesting property of these states is that they minimize
uncertainty in a well defined sense.  The product $\Delta q\Delta p$, where $q$ and $p$ are position and momentum, has no meaning in our setting because our variables have no natural ordering.
We therefore
express uncertainty in information-theoretic terms, specifically in terms of the R\'enyi
entropy of order 2 (which we call simply ``R\'enyi entropy" for short).  Moreover we 
consider not just the ``axis variables,'' but also variables associated with
all the other directions in the discrete phase space.  (In the continuous case these
other directions would be associated with linear combinations of $q$ and $p$.)  We will find that {\em each}
rotationally invariant state minimizes the R\'enyi entropy, averaged over all these
variables.  This will leave us with the question of picking out a ``most pointlike" of the
rotationally invariant states, if such a notion can be made meaningful; we 
address this question briefly in the conclusion.  

\section{DISCRETE PHASE SPACE }
\label{sec-2}

Over the years there have been many proposals for generalizing 
the Wigner function to discrete systems.  (See, for example, Refs.~\cite{BH,Marmo} and papers
cited in Ref.~[1].)  Here we adopt the discrete
Wigner function proposed by Gibbons {\em et al.} \cite{GHW}, which is well suited to
a system of qubits.  
The basic idea is to use, instead of the field of real numbers in which position and
momentum normally take their values, a {\em finite} field with a number of elements
equal to the dimension $d$ of the state space.  There exists
a field with $d$ elements if and only if $d$ is a power of a prime; so this approach
applies directly only to quantum systems, such as a collection of qubits, 
whose state-space dimension is such a number.

The two-element field ${\mathbb F}_2$ is simply the set $\{0,1\}$ with addition and multiplication 
mod 2, but the field of order $2^n$ with $n$ larger than 1 is different from arithmetic mod $2^n$.  For example,
${\mathbb F}_4$ consists of the elements $\{0,1,\omega,\omega+1\}$, in which $0$ and $1$ act as in 
${\mathbb F}_2$ and arithmetic involving the abstract symbol $\omega$ is determined by the equation 
$\omega^2 = \omega + 1$.    

The discrete phase space for a system of $n$ qubits is a two-dimensional vector
space over ${\mathbb F}_{2^n}$; that is, a point in the phase space can be expressed
as $(q,p)$, where $q$ and $p$, the discrete analogues of position and momentum, take values
in ${\mathbb F}_{2^n}$.  In this phase 
space it makes perfect sense to speak of lines and 
parallel lines; a line, for example, is the solution to a linear equation.  The key idea in constructing a Wigner function is to assign a pure quantum state,
represented by a one-dimensional
projection operator $Q(\lambda)$,
to each line $\lambda$ in phase space.  The only requirement imposed on the function 
$Q(\lambda)$ is
that it be ``translationally covariant."  This means that if we translate the line $\lambda$
in phase space by adding a fixed vector $(q,p)$ to each point, the associated quantum state changes
by a unitary operator $T_{(q,p)}$ associated with $(q,p)$.  The unitary translation operator $T_{(q,p)}$ is defined to be
\begin{equation}
T_{(q,p)} = X^{q_1}Z^{p_1} \otimes \cdots \otimes X^{q_n}Z^{p_n},
\end{equation}
where $X$ and $Z$ are Pauli operators
and
$q_i$ and $p_i$, which are elements of ${\mathbb F}_2$, 
are components of $q$ and $p$ when they are expanded
in particular ``bases'' for the field: e.g., $q = q_1b_1 + \cdots + q_nb_n$, where $(b_1,\ldots,b_n)$
is the basis chosen for the coordinate $q$.\footnote{The bases for $q$ and $p$ cannot be chosen
independently: each must be proportional to the {\em dual} of the other\cite{GHW}.}  One finds that the requirement of translational covariance
severely constrains the construction:
\begin{enumerate}
\item States assigned to parallel lines must be orthogonal.  A complete set of parallel lines, 
or ``striation,'' consists of exactly $d$ lines; so the states associated with a given striation
constitute a complete orthogonal basis for the state space.  In other words, each striation is
associated with a complete orthogonal measurement on the system.
\item The bases associated with different striations must be {\em mutually unbiased}.  That is,
each element of one basis is an equal-magnitude superposition of the elements of any of the
other bases.  There are exactly $d+1$ striations, so this construction generates a set of
$d+1$ mutually unbiased bases.  (Such a set is just sufficient for the complete tomographic
reconstruction of an unknown quantum state.)  
\end{enumerate}

Despite these constraints, there are 
many allowed
functions $Q(\lambda)$.  This implies that there are many possible definitions of the Wigner function for a system
of qubits, because 
once we have chosen a particular assignment of
quantum states to phase-space lines, the Wigner function of any quantum state
is uniquely fixed by the requirement that
the sums over the lines of any striation be equal to the probabilities of the outcomes of the
corresponding measurement.

\section{ROTATIONALLY INVARIANT STATES}

In the finite field, consider a quadratic polynomial $x^2 + ax + b$ that has no roots.  Then the equation
\begin{equation}
q^2 + aqp + bp^2 = c,
\end{equation}
with $c$ taking all nonzero values in ${\mathbb F}_{2^n}$, defines what we will call a set of 
``circles'' centered at the origin.  Fixing the values of $a$ and $b$---this is somewhat analogous to 
fixing the scales of the axes in the continuous case---we define a {\em rotation}
to be any linear transformation of the phase space that
leaves each circle invariant.\footnote{A different notion of rotation has been used 
in Ref.~\cite{Klimov}.}  (We consider only rotations around the origin.  A state centered
at the origin can always be translated to another point by $T_{(q,p)}$.)  For example, in the two-qubit phase space, our circles can be
defined by the equation
\begin{equation}
q^2 +  q p + \omega p^2 = c,
\end{equation}
and an example of a rotation is the transformation $R$ defined by 
\begin{equation}
\mtx{c}{q' \\ p'} = R\mtx{c}{q \\ p} = \mtx{cc}{1 & 1 \\ \omega+1 & \omega}\mtx{c}{q \\ p}.
\end{equation}
One can check that this particular rotation has the property that if we apply it 
repeatedly, starting with any
nonzero vector, it generates the entire circle on which that vector lies.  In this sense $R$
is a {\em primitive} rotation.  

With every 
unit-determinant linear transformation $L$
on the phase space, one can associate (though not uniquely) 
a unitary transformation $U$ on the state space whose action by conjugation on the translation
operators $T_{(q,p)}$ mimics the action of $L$ on the corresponding points of phase space\cite{Chau, GHW}.\footnote{The argument in Appendix B.3 of Ref.~\cite{GHW}
contains an error: 
Eqs.~(B24) and (B25) implicitly assume that the chosen field basis is self-dual, which is
not in fact the case.  However, the proof can be repaired by starting with a self-dual basis
to get those equations, and then changing to the actual basis via the argument 
of Appendix C.1.  That there exists a self-dual basis for ${\mathbb F}_{2^n}$ is
proved in Ref.~\cite{selfdual}.}  One can show that every rotation has unit determinant and must therefore
have an associated unitary transformation.  For example, for the rotation $R$ given above, 
if we expand both $q$ and $p$ in the field basis $(b_1,b_2) = (\omega, \omega+1)$, the 
following unitary transformation acts in the desired way on the translation operators:
\begin{equation}
U = \frac{1}{2}\mtx{cccc}{1 & i & i & -1 \\ i & 1 & -1 & i \\ 1 & i & -i& 1 \\ -i & -1 & -1 & i}.
\end{equation}
Thus just as
\begin{equation}
R\mtx{c}{1 \\ 0} = \mtx{cc}{1 & 1 \\ \omega+1 & \omega}\mtx{c}{1 \\ 0}
= \mtx{c}{1 \\ \omega + 1},
\end{equation}
we have that 
\begin{equation}
UT_{(1, 0)}U^\dag = U(X\otimes X)U^\dag = iX\otimes (XZ)  \, \propto T_{(1,\omega+1)}.
\end{equation}

For any number $n$ of qubits, let $R$ be a primitive rotation, and let $U$ be a unitary 
transformation associated with $R$ in the above sense.  
(Techniques for finding $U$ can be found in Refs.~\cite{GHW,Chau}.)  Then from the action of $U$ 
on the translation operators, it follows that $U$ acts in a particularly simple way on the
mutually unbiased bases associated with the striations of phase space: starting with any
one of these bases, repeated applications of $U$ generate all the other bases cyclically.  
That there always exists a unitary $U$ generating a complete set
of mutually unbiased bases for $n$ qubits 
has been shown by Chau \cite{Chau}.  In our present context,
we will reach the same conclusion by showing, in the following paragraph, 
that there always exists a primitive
rotation.  The existence of such a unitary matrix $U$ 
leads naturally to a simple prescription for choosing the function $Q(\lambda)$: 
(i) Use the translation operators to assign computational basis states 
to the
vertical lines.  (ii) Apply $U$ repeatedly to these states, and $R$ repeatedly 
to the lines, 
in order to complete the correspondence.
This prescription results in a definition of the
Wigner function that is ``rotationally covariant,'' in the sense that when one transforms the
density matrix by $U$, the values of the Wigner function are permuted among the
phase-space points according to~$R$.  

How does one find a primitive rotation $R$?  First, for any number of qubits, there always exists
a {\em primitive} polynomial of the form $x^2+x+b$ \cite{poly}, which one can use to define
circles by the equation $q^2 + qp + b p^2 = c$.
Then the linear transformation 
\begin{equation}
L = \mtx{cc}{1 & b \\ 1 & 0}
\label{L}
\end{equation}
is guaranteed to cycle through all the nonzero points of phase space \cite{Lidl}, and it 
always takes circles
to other circles.  Raising $L$ to the power $d-1$ gives us a unit-determinant transformation
that preserves circles and is indeed a primitive rotation.  
Moreover, one can write $R$ explicitly in terms of $b$:
\begin{equation}
R = L^{d-1} = \mtx{cc}{1 & 1 \\ b^{-1} & b^{-1}+1}.
\end{equation}

With $Q(\lambda)$ chosen in the way we have prescribed, the eigenstates of $U$ are 
our rotationally invariant states.  When we apply $U$ to {\em any} state, the Wigner function
simply flows along the circles in accordance with the rotation $R$.  But an eigenstate of $U$
does not change under this action, so its Wigner function must be constant on each circle.  

\section{MINIMIZING ENTROPY}

Consider again our complete set of $d+1$ mutually unbiased bases, and let $|ij\rangle$ be the
$j$th vector in the $i$th basis.  These vectors together have the following 
remarkable property: for any pure state $|\psi\rangle$, the probabilities $p_{ij} = |\langle \psi|ij\rangle|^2$
satisfy \cite{2design, 2design2}
\begin{equation}
\sum_{ij} p_{ij}^2 = 2.
\end{equation}
Now consider the R\'enyi entropy $H_R = -\log_2\left(\sum_j p_{ij}^2\right)$ of the 
outcome-probabilities  of the
$i$th measurement when applied to the state $|\psi\rangle$.  This entropy is a measure of our 
inability to predict the outcome of the measurement.  The {\em average}
of $H_R$ over all the mutually unbiased measurements can be bounded from below \cite{Ballester}:
\begin{equation}
\langle H_R \rangle = \left(\frac{1}{d+1}\right)\sum_i\left[-\log_2\left(\sum_j p_{ij}^2\right)\right]
\geq -\log_2\left[\left(\frac{1}{d+1}\right)\sum_{ij} p_{ij}^2\right] = \log_2(d+1) - 1,
\label{ineq}
\end{equation}
with equality holding only if the R\'enyi entropy is {\em constant} over all the mutually unbiased
measurements.\footnote{The analogous inequality in terms of Shannon entropy was
proved in Refs.~\cite{uncertainty, uncertainty2}.}

Now, for any of the rotationally invariant states defined in the last section, the R\'enyi entropies
associated with the $d+1$ mutually unbiased measurements are indeed equal.  By the inequality (\ref{ineq}), such states therefore minimize the average
R\'enyi entropy over all these measurements, that is, over all the directions in phase space.  

\section{EXAMPLES}

The one-qubit case is very simple.  The three mutually unbiased bases generated in our construction
are the eigenstates of the Pauli operators $X$, $Y$, and $Z$.  It is not hard to find a unitary
transformation that cycles through these three bases.  Such a
transformation rotates the Bloch sphere by $120^\circ$ around the axis $(x,y,z) = (1,1,1)$.  The two
eigenstates of this unitary transformation, which are the eigenstates of $X+Y+Z$,
are rotationally invariant: each of their Wigner functions is constant on the only circle in the
$2\times 2$ phase space.  And each of these states minimizes the average R\'enyi entropy
for the measurements $X$, $Y$, and $Z$.  It is interesting to note that one of these two 
states has a positive Wigner function.  

Clearly there is nothing intrinsically special about these two states.  They are special only in relation
to the three measurements $X$, $Y$, and $Z$, which are associated with the three striations
of the phase space.  But in the context of quantum cryptography, 
the entropy-minimization property is quite relevant.  In the six-state scheme (in 
which the signal states are the eigenstates of $X$, $Y$, and $Z$), if Eve chooses to eavesdrop by making a
complete measurement on certain photons, her best choice is to make a measurement whose outcome-states
are entropy-minimizing in our sense: it turns out that such a choice minimizes
Eve's own R\'enyi entropy about Alice's bit.

An interesting example comes from the 3-qubit case.  The relevant field is ${\mathbb F}_8$, which 
can be constructed from ${\mathbb F}_2$ by introducing an element $b$ that is defined to satisfy
the equation $b^3 + b^2 + 1 = 0$.  In our $8 \times 8$ discrete phase space, we can define circles
via the equation
\begin{equation}
q^2 + qp + p^2 = c,
\label{3circ}
\end{equation}
where $c$ can take any nonzero value.  A primitive rotation preserving these circles is\footnote{Even
though Eq.~(\ref{3circ}) is not of the form we used in reaching Eq.~(\ref{L}), in that it is not based on
a primitive polynomial, the matrix $R$ is nevertheless a primitive rotation.}
\begin{equation}
R = \mtx{cc}{b^3 & b^6 \\ b^6 & b^5}.
\end{equation}

One finds that of the eight eigenvectors of any unitary $U$ corresponding to $R$, all
of which are rotationally invariant, exactly
one has a positive Wigner function for a specific, fixed function $Q(\lambda)$ associated with $U$.  
This state is also easy to describe physically.
For a particular choice of $U$, it is of the form
\begin{equation}
|\psi\rangle = \sqrt{1/3}|+++\rangle + \sqrt{2/3}|---\rangle,
\end{equation}
where $|+\rangle$ and $|-\rangle$ are the two eigenstates (with a specific relative phase)
of the operator
$X+Y+Z$.  
If we regard $|\psi\rangle$ as analogous to a coherent
state at the origin, then
the coherent-like states at the 63 other
phase-space points can be obtained from $|\psi\rangle$ by applying Pauli rotations to the individual qubits.  The Wigner function of each of these states has the value 0.319 at its center, the largest
value possible for any three-qubit state.  

\section{CONCLUSION}

We have found that one can make sense of the notion of rotational invariance in 
a discrete phase space for a system of $n$ qubits.  The rotationally invariant states
are in this respect analogous to the energy eigenstates of a harmonic oscillator,
but the analogy is not perfect.  Our rotationally invariant states are
all states of minimum uncertainty with respect to
the various directions in phase space, whereas except for the ground state, the harmonic oscillator eigenstates 
do not have this property (the uncertainty, even in our R\'enyi sense, increases 
with increasing energy).  We have considered the further restriction to positive 
Wigner functions but so far have found examples of such states only for a single qubit and
for three qubits.  However, for any number of qubits,
one can show that at least one of our rotationally invariant states takes a value at its
center equal to the maximum value attainable by the Wigner function 
of {\em any} state.  Perhaps this latter property,
rather than positivity, should be taken as the defining feature of a ``most pointlike" state.


\end{document}